\documentstyle[twoside,fleqn,espcrc2,epsf]{article}

\newcommand{\AmS}{{\protect\the\textfont2
  A\kern-.1667em\lower.5ex\hbox{M}\kern-.125emS}}

\title{Progress report on hadron spectroscopy with improved actions
\thanks{
Poster by Artan Bori\c{c}i at LATTICE96.}}

\author
{Artan Bori\c{c}i and Philippe de Forcrand
\address{CSCS/SCSC \\
         ETH-Zentrum, RZ \\
         CH-8092 Z\"urich}}

\begin{document}

\begin{abstract}
Quenched light hadron masses are measured on blocked lattices, using 6
different lattice discretizations of the Dirac operator. Results are compared
with those of unblocked lattices, allowing for a ``ranking'' of the Dirac
discretizations.
\end{abstract}

\maketitle

\section{INTRODUCTION}

Two years ago we reported on tadpole-improved fermion actions
on blocked lattices \cite{us_lat94}. We showed that tadpole
improvement was necessary for the clover improvement to give accurate
results (see also \cite{Lepage_lat95}).
Since then many groups have tested the same action and confirmed this
conclusion \cite{spectrum}.

Here, we include other improved fermion actions with and without tadpole
improvement. The aim is to assess the importance of tadpole renormalizations
for those actions and to identify the ``best'' discretization(s)
of the Dirac operator.

We consider improvements of the {\bf Wilson fermion action} \cite{Wilson2}
%
%
which has $O(a)$ leading discretization errors.
For improvements of the pure-gauge Wilson action we refer to
\cite{Lepage_lat95}. They are less important since the discretization
error is $O(a^2)$ in the gauge sector. We use quenched blocked configurations
of \cite{Taro}
for which the discretization errors are even smaller. By measuring the
string tension on the blocked lattices,
we observed the restoration of the rotational symmetry. For further details
see \cite{Takaishi}.

One of the first proposals for improved fermion actions was the
{\bf Eguchi-Kawamoto action} \cite{Egu_Kawa}:
\begin{equation}
\begin{array}{l}
D^{(EK)}_{ij} = \delta_{ij} - \kappa ~\sum_{\mu = 1}^{4}
~[(r - \gamma_{\mu} ) U_{\mu i} \delta_{i + \hat{\mu} , j} \\
+ (r + \gamma_{\mu} ) U_{\mu i - \hat{\mu}}^{\dagger} \delta_{i - \hat{\mu} , j} ] \\
+ \frac{\kappa}{8} ~\sum_{\mu = 1}^{4}
~[(r' - \gamma_{\mu} ) U_{\mu i} U_{\mu i +2\hat{\mu}} \delta_{i + 2\hat{\mu} , j} \\
+ (r' + \gamma_{\mu} ) U_{\mu i -  \hat{\mu}}^{\dagger}
U_{\mu i - 2\hat{\mu}}^{\dagger}\delta_{i - 2\hat{\mu} , j}]
\end{array}
\end{equation}
with $r' = 2r$ and $\kappa_c = 1/6$. We set here $r = 1$.
It cancels the $O(a^2)$ errors of the naive discretization
and breaks the chiral symmetry by $O(a^3)$ terms. Therefore,
it has $O(a^3)$ discretization errors.
Computations with this action can be found also in \cite{FiebWolosh}.

The {\bf Clover-improved Wilson action}
is constructed to remove the $O(a)$ errors of the Wilson action by adding
the term \cite{Sheik}:
\begin{equation}\label{clover_term}
\Delta W^{(clover)} = c \frac{i\kappa}{2} \sum_{\mu\nu i} \sigma_{\mu\nu}
{\cal P}_{\mu\nu,i}
\end{equation}
with $c = 1$, $\sigma_{\mu\nu} = \frac{i}{2} [\gamma_{\mu}, \gamma_{\nu}]$
and ${\cal P}_{\mu\nu,i}$ the antisymmetric and antihermitian lattice
operator that discretizes the field strength tensor $F_{\mu\nu}$ by
the nearest-neighbor plaquette terms
at lattice site $i$ in the $\mu\nu$ plane.
$\Delta W^{(clover)}$ has $O(a)$ discretization errors.

Note that the clover action can be obtained from the
Eguchi-Kawamoto action by the
isospectral transformation \cite{Mrtnl_Sachr_Vlad}
\begin{equation}
\begin{array}{l}
\psi \rightarrow \psi - \frac{a}{4} D\psi \\
\bar{\psi} \rightarrow \bar{\psi} + \frac{a}{4} \bar{\psi}D^{\dag}
\end{array}
\end{equation}
after dropping $O(a^2)$ terms (and higher) and proper normalization, where
%
%
$D$ is the (massless) naive lattice Dirac operator.

\subsection{Tadpole improvement}

All the actions above have additional $O(g^2a)$ errors at the one-loop level in
perturbation theory (see for example \cite{Heatl_Schr_Mrtnl_Pitt_Ross}).
These errors cannot be removed perturbatively for the usual lattice cutoffs
of the order 1 GeV.
Therefore, non-perturbative treatments of this problem have recently been
proposed.

Tadpole improvement is a mean field prescription which rescales gauge fields
$U_{\mu,i} \rightarrow U_{\mu,i} / u_0$ with $u_0 = P^{1/4}$,
$P$ being the average plaquette \cite{Lepage_Mackenzie}.
Another alternative is to tune the clover coefficient $c$ non-perturbatively
so that $O(a)$ errors disappear \cite{Luesch_Sint_Somm_Weisz}.
However, tadpole improvement
is trivial to implement and cancels most of the $O(a)$ errors in our application.

Another improved fermion lattice action is the D234 action
\cite{Alford_Klassen_Lepage}.
It can be written in the form
\begin{equation}
D234 = D^{EK}(r'=r=1) + \frac{1}{2} \Delta W^{(clover)}
\end{equation}
with tadpole improved gauge fields.
Like the EK action it has $O(a^3)$ classical discretization errors.
The free $D234$ action has $\kappa_c = 1/7$.

\section{LIGHT HADRON SPECTRUM WITH IMPROVED ACTIONS}

We have calculated the light hadron masses in quenched QCD for the
following fermion actions \cite{thesis}:
\[
\begin{array}{ll}
1)  &  Wilson ~(W)\\
2)  &  Wilson-Clover ~(C)\\
3)  &  Wilson-Clover-Tadpole ~(C_{LM})\\
4)  &  Eguchi-Kawamoto ~(EK)\\
5)  &  Eguchi-Kawamoto-Tadpole ~(EK_{LM})\\
6)  &  D234 \\
\end{array}
\]
on $8^3\times16$ configurations
obtained from $32^3\times64$ lattices at $\beta =6$ after 2 blocking steps.
We compare our results with those
on the original unblocked lattice with Wilson fermions
taken from \cite{Gupta}; we denote them by $G$.
In all cases a sample of 100 configurations is considered.

The string tension measured on blocked lattices is
$a'^2 K = 0.93(5)$. Comparison with the experimental
value $K = (440$ MeV $)^2$ gives
$a'^{-1} = 456(12)$ MeV or
$a' = 0.432(11)$ fm.
By matching this string tension
with that of the Wilson gauge action,
we found that $a' = 0.432(11)$ fm corresponds to $\beta \approx 5.1$.
The string tension on the fine lattice is taken from \cite{Bali_Schilling},
$a^2 K = 0.0513(25)$ corresponding to $a = 0.101(2)$ fm.

In Table 1 we compare the additive quark mass renormalization
for different quark actions at $\kappa = \kappa_c$, i.e.
$\Delta m = 1/2(1/\kappa_c - 1/\kappa_c^{free})$ for
$W, C, C_{LM}$ and
$\Delta m = 2/3(1/\kappa_c - 1/\kappa_c^{free})$ for
$EK, EK_{LM}, D234$, where
$\kappa_c^{free}$ is the free $\kappa_c$.

\vspace{.2cm}
Table 1

\begin{center}
\begin{tabular}{|l|c|}
\hline\hline
Action & $\Delta m$ \\
\hline
$W$       &  -1.50(2)  \\
$C$       &  -1.06(2)  \\
$C_{LM}$  &   0.17(1)  \\
$EK$      &  -1.10(4)  \\
$EK_{LM}$ &  -0.49(1)  \\
$D234$    &  -0.222(4)  \\
\hline\hline
\end{tabular}
\end{center}
\vspace{.3cm}

In Table 2 we compare the $\rho, N$ and $\Delta$ masses
extrapolated to the chiral limit
by using the quadratic fit: $m_0 + b (a m_{\pi})^2$.
We see that the results on the coarse lattices
approach monotonically those on the fine lattices (with the Wilson action)
if we order the improved quark actions as:
$W, C, EK, EK_{LM}, D234, C_{LM}$.

\vspace{.2cm}
Table 2

\begin{center}
\begin{tabular}{|l|l|l|l|}
\hline\hline
Action & $m_{\rho}/\sqrt{K}$ & $m_N/\sqrt{K}$ & $m_{\Delta}/\sqrt{K}$\\
\hline
$W$       &  0.95(1)  &  1.79(3)  &  1.93(1)  \\
$C$       &  1.130(2) &  1.85(2)  &  2.12(2)  \\
$C_{LM}$  &  1.42(1)  &  2.18(4)  &  2.53(2)  \\
$EK$      &  1.22(1)  &  1.89(5)  &  2.19(1)  \\
$EK_{LM}$ &  1.35(1)  &  1.92(6)  &  2.37(1)  \\
$D234$    &  1.39(1)  &  2.03(5)  &  2.48(1)  \\
\hline
$G$       &  1.48(1)  &  2.17(2)  &  2.56(1)  \\
\hline\hline
\end{tabular}
\end{center}
\vspace{.3cm}

In Table 3 the hadron masses are extrapolated by adding a cubic term in the fit.
In this case the above ordering remains, except the $D234$ and $C_{LM}$ actions
exchange places.

\vspace{.2cm}
Table 3

\begin{center}
\begin{tabular}{|l|l|l|l|}
\hline\hline
Action & $m_{\rho}/\sqrt{K}$ & $m_N/\sqrt{K}$ & $m_{\Delta}/\sqrt{K}$\\
\hline
$W$       &  0.883(3) &  1.59(2)  &  1.82(2)  \\
$C$       &  1.12(1)  &  1.64(1)  &  1.96(2)  \\
$C_{LM}$  &  1.42(6)  &  1.75(5)  &  2.38(2)  \\
$EK$      &  1.22(3)  &  1.652(1) &  2.24(3)  \\
$EK_{LM}$ &  1.30(4)  &  1.73(23) &  2.3255   \\
$D234$    &  1.41(2)  &  1.82(4)  &  2.41(4)  \\
\hline
$G$       &  1.41(1)  &  2.05(2)  &  2.62(1)  \\
\hline\hline
\end{tabular}
\end{center}
\vspace{.3cm}

\begin{figure}[htb]
\centerline{\epsfysize = 4 in \epsffile {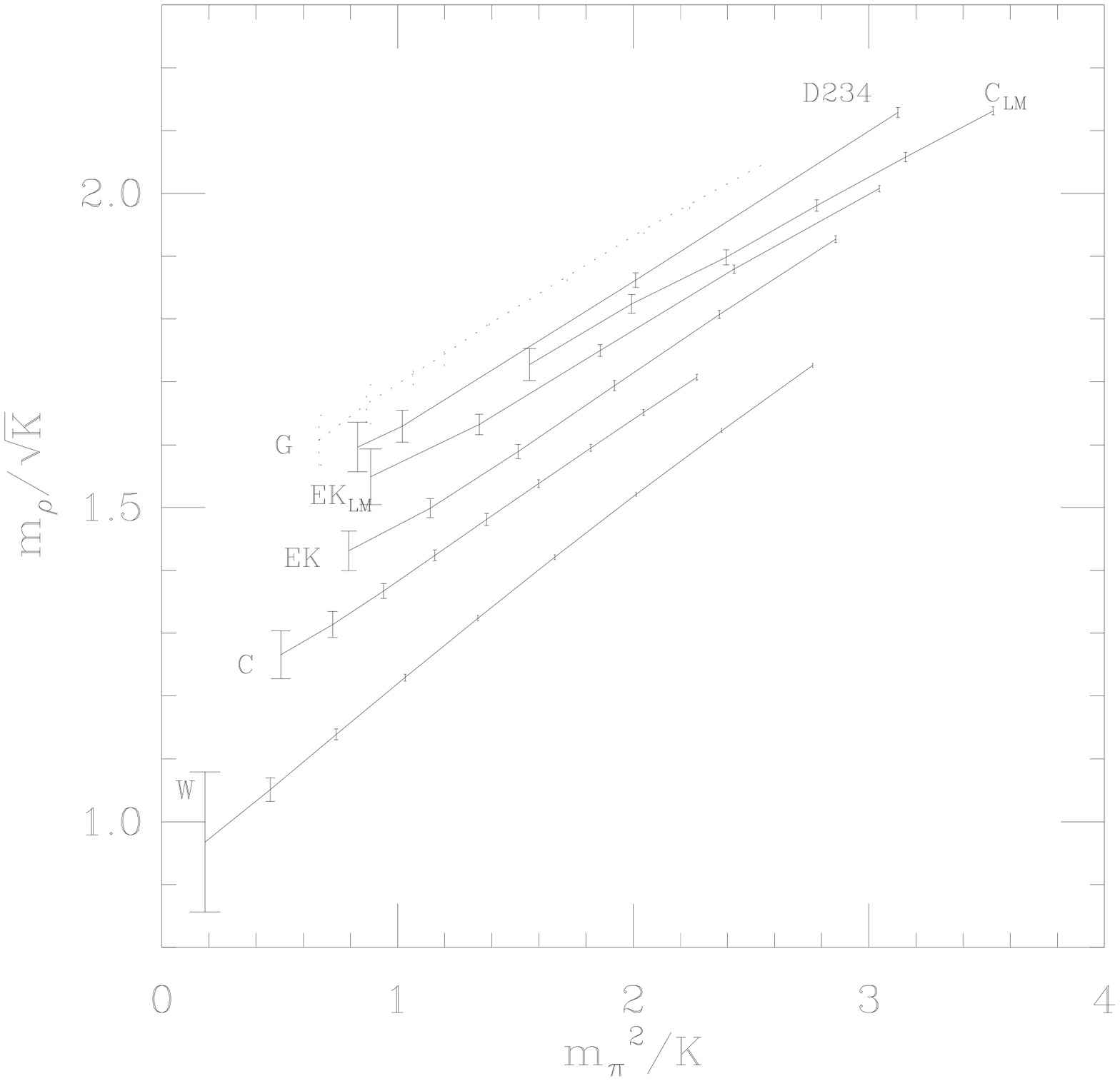}}
\vspace{-4cm}
\caption{
Comparison of the $\rho$ masses for different quark actions
indicated in the figure.
The dotted line was obtained with
Wilson fermions on a fine lattice ($\beta = 6$, $32^3\times64$) \protect\cite{Gupta}.
The rest of the results
comes from a twice-blocked lattice ($\beta \approx 5.1$, $8^3\times16$).
}
\end{figure}

\begin{figure}[htb]
\centerline{\epsfysize = 4 in \epsffile {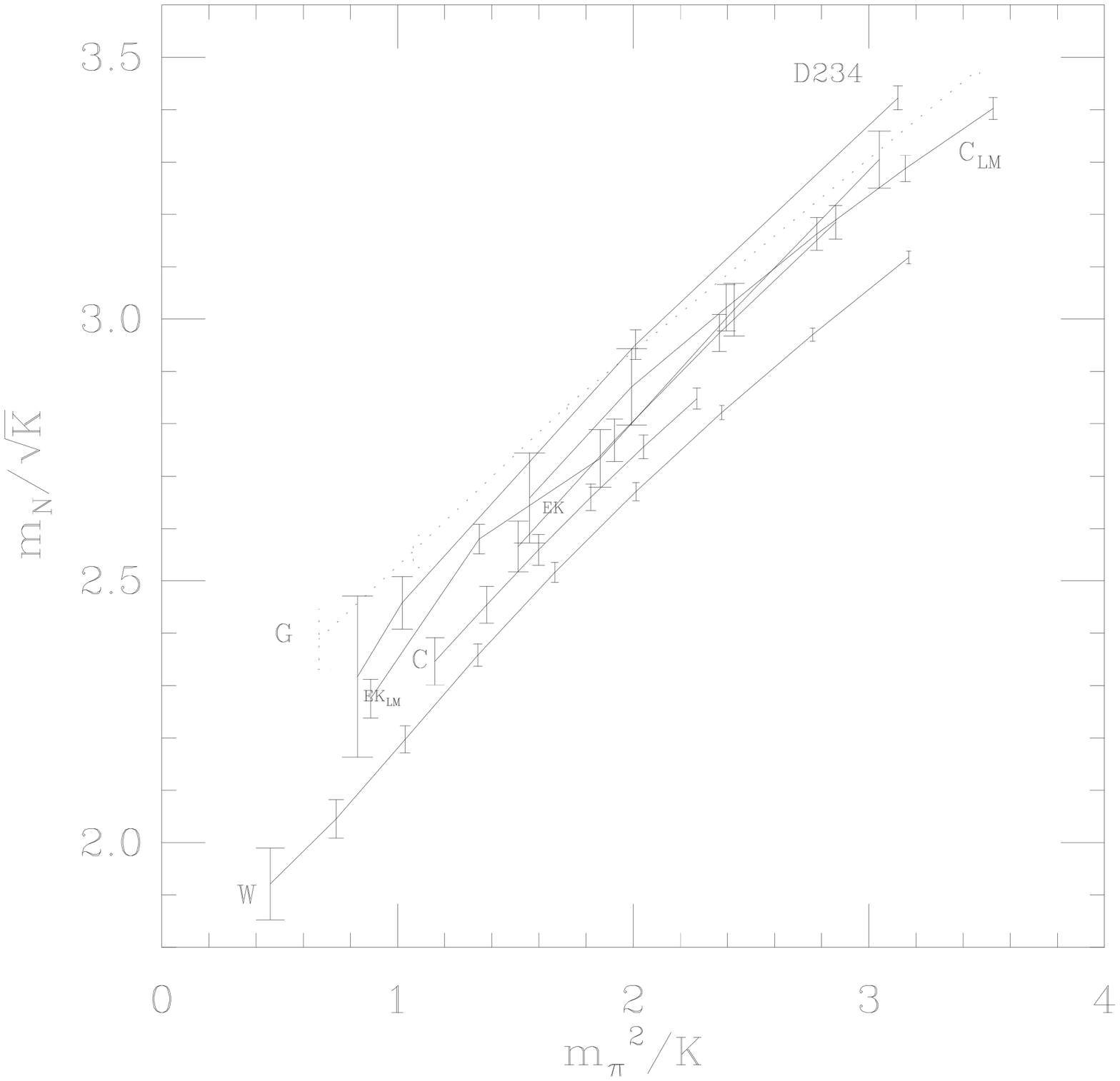}}
\vspace{-4cm}
\caption{
Comparison of the Nucleon masses for different quark actions
indicated in the figure.
The dotted line was obtained with
Wilson fermions on a fine lattice ($\beta = 6$, $32^3\times64$) \protect\cite{Gupta}.
The rest of the results
comes from a twice-blocked lattice ($\beta \approx 5.1$, $8^3\times16$).
}
\end{figure}

\begin{figure}[htb]
\centerline{\epsfysize = 4 in \epsffile {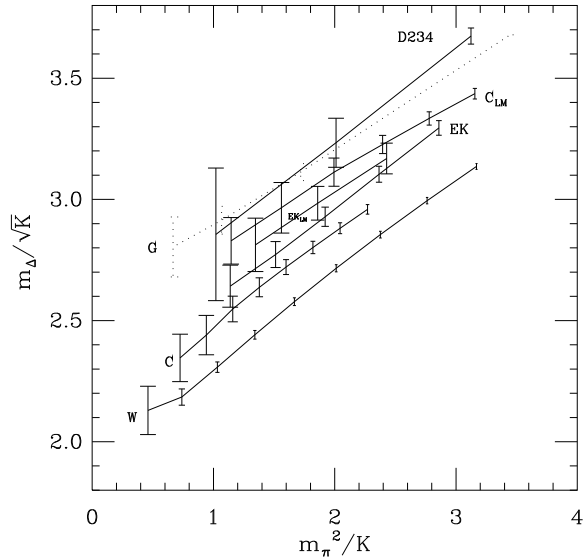}}
\vspace{-4cm}
\caption{
Comparison of the $\Delta$ masses for different quark actions
indicated in the figure.
The dotted line was obtained with
Wilson fermions on a fine lattice ($\beta = 6$, $32^3\times64$) \protect\cite{Gupta}.
The rest of the results
comes from a twice-blocked lattice ($\beta \approx 5.1$, $8^3\times16$).
}
\end{figure}

These results confirm our earlier calculations that tadpole improvement
is crucial to approach the continuum limit faster \cite{us_lat94}. On the
other hand tadpole effects for the $EK$ action are about half of those
for the clover action. This is consistent with the expectation
that the higher the order of gauge
fields in a given operator, the higher the tadpole renormalizations.
The $EK$ action will be therefore less sensitive to the precise value
of the tadpole coefficient $u_0$.

In Figures 1, 2, 3 we compare the $\rho, N$ and $\Delta$ masses
respectively for the different fermionic actions.
All of them are compared
to the results obtained by the Wilson action on the fine lattice.
They show that the $D234$ action approaches better the results
of fine lattices.
In Figure 4 we show an Edinburgh plot; it favors the tadpole-improved 
clover action.


To summarize, we find that the tadpole-improved clover action,
the tadpole-improved Eguchi-Kawamoto action and the $D234$
action are much better discretizations of the
continuum

\clearpage

\vspace{-2cm}
\begin{figure}[htb]
\centerline{\epsfysize = 4 in \epsffile {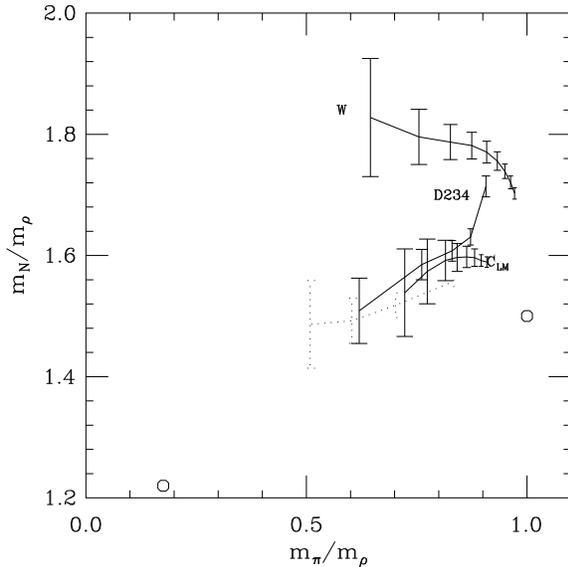}}
\vspace{-4cm}
\caption{
Edinburgh plot for the Wilson and the two most improved actions,
$D234$ and $C_{LM}$.
The dotted line was obtained with
Wilson fermions on a fine lattice ($\beta = 6$, $32^3\times64$) \protect\cite{Gupta}.
The rest of the results
comes from a twice-blocked lattice ($\beta \approx 5.1$, $8^3\times16$).
The octagons are the physical ratio and infinite quark mass limit.
}
\end{figure}

\noindent
Dirac operator than the Wilson action, allowing for
an increase of the lattice spacing by a factor $\sim 4$.
Among these 3 actions, the greater simplicity and the higher locality
of the tadpole-improved clover action make it the most appealing.

\end{document}